\documentclass[conference]{IEEEtran}
\IEEEoverridecommandlockouts
\usepackage{cite}
\usepackage{url}
\usepackage{amsmath,amssymb,amsfonts}
\usepackage{algorithmic}
\usepackage{algorithm}

\usepackage{graphicx}
\usepackage{textcomp}
\usepackage{xcolor}
\usepackage[nolist]{acronym}
\def\BibTeX{{\rm B\kern-.05em{\sc i\kern-.025em b}\kern-.08em
    T\kern-.1667em\lower.7ex\hbox{E}\kern-.125emX}}

\usepackage{BNPackages/BNPackages}
\usepackage{balance}
\usepackage{BNPackages/BNMathMacros}
\usepackage{BNPackages/BNPlots}
\usepackage{BNPackages/BNMathEstimation}

\pgfplotsset{BNplotstyle}
\newlength\fheight
\newlength\fwidth
\newlength\plotheight
\newlength\figheight
\newlength\figstepheight
\usepackage[font=small,skip=0pt, labelformat=simple]{subcaption}

\usetikzlibrary{calc}
\usetikzlibrary{positioning}
\usetikzlibrary{external}
\usetikzlibrary{spy}

\DeclareMathAlphabet\mathbfcal{OMS}{cmsy}{b}{n}

\RenewDocumentCommand{\meas} {s o d()} {%
	\ensuremath{%
		\IfBooleanTF{#1}{\rv}{\rvec}y
		\IfNoValueF{#2}{_{#2}}
		\IfNoValueF{#3}{^{#3}}
	}%
}

\NewDocumentCommand{\triggermat} {t~ o d()} {%
	\ensuremath{%
	\IfBooleanT{#1}{\bigl(}
	\mat{Z}%
	\IfNoValueF{#2}{_{#2}}
	\IfNoValueF{#3}{^{#3}}
	\IfBooleanT{#1}{\bigr)}
	}%
}

\NewDocumentCommand{\triggerconst} {o d()} {%
	\ensuremath{%
		\vec{c}%
		\IfNoValueF{#1}{_{#1}}
		\IfNoValueF{#2}{^{#2}}
	}%
}

\NewDocumentCommand{\sysmult} {t~ o d()} {%
	\ensuremath{%
		\IfBooleanT{#1}{\bigl(\!}
		\mathbfcal{A}
		\IfNoValueF{#2}{_{#2}}
		\IfNoValueF{#3}{^{#3}}
		\IfBooleanT{#1}{\bigr)}
	}%
}

\NewDocumentCommand{\measmult} {o d()} {%
	\ensuremath{%
		\mathbfcal{H}
		\IfNoValueF{#1}{_{#1}}
		\IfNoValueF{#2}{^{#2}}
	}%
}

\DeclareMathOperator*{\Exp}{\mathbb{E}}

\def\Width{0\kern0.5\tabcolsep\dots\tabcolsep0}
\def\Widtha{0\kern0.5\tabcolsep\dots\kern1cc\tabcolsep0}

\theoremstyle{definition}
\newtheorem{remark}{Remark}

\begin{document}
\begin{acronym}
    \acro{KF}[KF]{Kalman filter}
    \acro{SEBKF}{stochastic event-based Kalman filter}
    \acro{CNKF}[CNKF]{colored-noise Kalman filter}
    \acro{FIR}{finite impulse response}
    \acro{ABC}{approximate Bayesian computation}
    \acro{MSE}[MSE]{mean squared error}
    \acro{MMSE}{minimum mean squared error}
    \acro{NEES}{normalized estimation error squared}
    \acro{ANEES}{average normalized estimation error squared}
    \acro{ML}{maximum likelihood}
    \acro{CI}{covariance intersection}
    \acro{FCI}{fast covariance intersection}
    \acro{EBFCI}{event-based fast covariance intersection}
    \acro{BSC}{Bar-Shalom-Campo}
    \acro{WLS}{weighted least squares}
    \acro{SSODP}{stochastic send-on-delta with prediction}
    \acro{SSOD}{stochastic send-on-delta}
    \acro{SOD}{send-on-delta}
    \acro{SODP}{send-on-delta with prediction}
    \acro{DARE}{discrete algebraic Riccati equation}
    \acro{AS}{augmented state}
    \acro{EBAS}{event-based augmented state}
    \acro{SSM}{state space model}
    \acro{AWGN}{additive white Gaussian noise}
    \acro{UT}{unscented transformation}
    \acro{UKF}{unscented Kalman filter}
    \acro{GND}{Generalized Normal Distribution}
    \acro{pdf}{probability density function}
    \acro{LTI}{linear time-invariant}
\end{acronym}

\title{A Unified Framework for Innovation-based Stochastic and Deterministic Event Triggers

}

\author{\IEEEauthorblockN{ Eva Julia Schmitt and Benjamin Noack}
\IEEEauthorblockA{\textit{Autonomous Multisensor Systems Group} \\
\textit{Institute for Intelligent Cooperating Systems}\\
\textit{Otto von Guericke University Magdeburg, Germany}\\
eva.schmitt@ovgu.de, benjamin.noack@ieee.org} }  


\maketitle

\begin{abstract}
Resources such as bandwidth and energy are limited in many wireless communications use cases, especially when large numbers of sensors and fusion centers need to exchange information frequently. One opportunity to overcome resource constraints is the use of event-based transmissions and estimation to transmit only information that contributes significantly to the reconstruction of the system's state. The design of efficient triggering policies and estimators is crucial for successful event-based transmissions. While previously deterministic and stochastic event triggering policies have been treated separately, this paper unifies the two approaches and gives insights into the design of reliable trigger-matching estimators. Two different estimators are presented, and different pairs of triggers and estimators are evaluated through simulation studies.
\end{abstract}

\begin{IEEEkeywords}
stochastic event-based triggers, deterministic triggers, event-based estimation 
\end{IEEEkeywords}

\section{Introduction}
\acresetall
In modern automated systems as smart cities, smart manufacturing, or smart farming large amounts of data need to be disseminated between spatially distributed nodes, e.\,g., sensors, fusion centers, and agents~\cite{Dargie2010, WSNSurvey}. Transmitting data wirelessly at high data rates requires large amounts of bandwidth and energy, which can both be limited resources. To reduce the burden on the communications systems in remote estimation problems, countermeasures need to be taken. The problem can be approached by reducing the amount of transmitted data through quantification of the information contained in every potential message and only transmitting if necessary or useful. This can be achieved with the help of event-based transmissions and estimation and usually leads to a trade-off between communication frequency and estimation error at the remote estimator~\cite{wuEventBasedSensorData2013}. Therefore, it is important to define adequate triggering policies and to find estimators that ensure consistency to guarantee reliable estimation results~\cite{molinConsistencyPreservingEventTriggeredEstimation2015}. Estimator consistency ensures that the estimated uncertainty matches the estimation error of the filter~\cite{julierNondivergentEstimationAlgorithm1997}. Furthermore, the use of estimators that can exploit the information contained in the non-fulfillment of the event triggering condition in between transmissions is desirable~\cite{sijsEventbasedStateEstimation2013}.

In the last years, two main directions have been followed in terms of event-based triggering: Deterministic policies that certainly trigger a transmission if a specified condition is met opposed to stochastic policies that trigger with a certain probability depending on the system's state. 
Deterministic policies include the \ac{SOD} scheme~\cite{heemelsAsynchronousMeasurementControl1999, miskowiczSendOnDeltaConceptEventBased2006}, \ac{SODP}~\cite{CRC15_Sijs} derived from this, matched sampling~\cite{marckRelevantSamplingApplied2010}, variance-based sampling~\cite{trimpeEventBasedStateEstimation2014}, and many other policies specifically designed for certain use cases. The most common stochastic policies are the stochastic versions of \ac{SOD} and \ac{SODP} and modifications thereof~\cite{hanStochasticEventTriggeredSensor2015,andrenEventBasedStateEstimation2016, schmittGaussianityPreservingEventBasedState2019}. All of these stochastic policies use Gaussian error weighting functions, which will be further explored in the following.

Both approaches have their advantages and disadvantages: 
Under deterministic policies it is ensured that a predefined error will not be exceeded before the next transmission is triggered and the achievable performance in terms of transmission rate versus estimation error is better than under stochastic policies~\cite{yuStochasticDeterministicEventBased2021}. However, this comes at the cost of a more complicated estimator design as the deterministic policy adds a non-Gaussian noise source to the system's measurement equation in non-transmission instants. Stochastic policies using Gaussian error weighting functions on the other hand provide a simple estimator design but degrade the system performance due to the introduced slack in the stochastic triggering decisions.
Different estimators that can handle deterministic or stochastic transmission policies have been developed, including a set-membership-based approach~\cite{CRC15_Sijs, CDC12_Noack} and event-based particle filters~\cite{ruuskanenInnovationBasedTriggeringEventBased2020,gasmiEventTriggeredStateEstimation2022, gasmiNonlinearEventbasedState2024} for deterministic policies and for stochastic policies extensions of the linear \ac{KF} for the standard case~\cite{hanStochasticEventTriggeredSensor2015,andrenEventBasedStateEstimation2016, schmittGaussianityPreservingEventBasedState2019} and for correlated input data~\cite{MFI23_Schmitt, Fusion24_Schmitt}.
Since stochastic policies and the associated estimators usually rely on such,
in this work we focus on innovation-based stochastic and deterministic triggering policies, where the error used in the triggering policy has a similar form as the innovation term used in Kalman filtering and is based on the difference between the current measurement and an arbitrary prediction of this measurement.

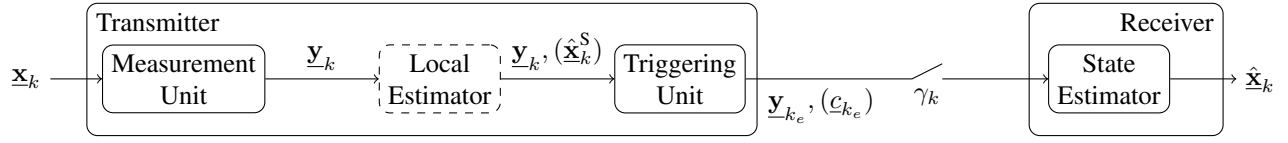
\begin{figure*}[t]
    \centering
     
\def\a{1.5}
\begin{tikzpicture}
    \draw (0,0) node[draw, align=center, anchor=west, rounded corners] (meas) {Measurement\\ Unit};
    \draw (meas.east) ++(\a,0) node[draw, align=center, anchor=west, rounded corners, dashed] (est) {Local\\ Estimator};
    \draw (est.east) ++(\a,0) node[draw, align=center, anchor=west, rounded corners] (trig) {Triggering\\ Unit};
    \draw (meas.west) ++(-1,0) node (x) {\(\state[k]\)};
    \draw (trig.east) ++(4, 0) node[draw, align=center, anchor=west, rounded corners] (estRec) {State\\ Estimator};
    \draw (estRec) ++(2, 0) node (xhat) {\(\est[k]\)};
    \draw (trig.east) ++(2.4, 0) node[below] (switch){\(\gamma_k\)};
    \draw (trig.east) ++(1.0, 0) node[below] (xe) {\(\meas[k_e], (\triggerconst_{k_e})\)};
    
    \draw[->] (meas) -- (est) node[midway, above] (y) {\(\meas[k]\)};
    \draw[->] (est) -- (trig) node[midway, above] (xhatS) {\(\meas[k], (\est_k^\text{S})\)};
    \draw[->] (x) -- (meas.west);
    \draw[-] (trig.east) -- ++(2.2, 0) -- ++(0.4, 0.2);
    \draw[->] (trig.east) ++(2.6,0) -- (estRec);
    \draw[->] (estRec.east) -- (xhat);
    
    \draw (meas.west) ++(-0.2,1) coordinate (box2ll);
    \draw (trig.east) ++(0.15,-0.75) coordinate (box2ur);
    \draw[rounded corners] (box2ll) rectangle (box2ur);
    \draw (box2ll) node[anchor = north west] {Transmitter};

    \draw (estRec.west) ++(-0.25,-0.75) coordinate (rbox2ll);
    \draw (estRec.east) ++(0.7,1.0) coordinate (rbox2ur);
    \draw[rounded corners] (rbox2ll) rectangle (rbox2ur);
    \draw (rbox2ur) node[anchor = north east] {Receiver};
\end{tikzpicture}
    \caption{System setup consisting of transmitter and receiver with event-based transmissions of \(\meas[k]\). The transmitter includes a measurement unit, a triggering unit and an optional local estimator which estimates \(\est[k]^\text{S}\) depending on the choice of \(\triggerconst[k]\).}
    \label{fig:system}
\end{figure*}

This work unifies the innovation-based deterministic and stochastic event triggering approaches in terms of trigger notion and design as well as it provides insights into suitable estimators. In particular, it is investigated under which conditions the more resource-efficient deterministic transmission policies can be combined safely with the simple stochastic estimator design to combine the advantages of both schemes.
The contributions can be summarized as follows. Firstly, a generalized framework for  stochastic triggers that embeds innovation-based deterministic triggers is proposed. It allows for better understanding of the connections between classical stochastic and deterministic event triggers and the derivation of design rules. 
Based on the new insights regarding the properties of different triggering policies, the consistency of the available stochastic event-based \ac{KF}~\cite{hanStochasticEventTriggeredSensor2015} with those triggering policies is investigated. Furthermore, an extension of the event-based particle filter~\cite{gasmiEventTriggeredStateEstimation2022} to stochastic policies is provided and improvements for the use with linear system models and deterministic policies are made. Finally, different combinations of the triggering policies and the two estimators are evaluated regarding the estimates' \ac{MSE} and consistency in simulation studies. 

\section{Notation}
An underlined variable $\vec{x}\in\IR^q$ denotes a real-valued vector. Lowercase boldface letters $\rvec{x}$ are used for quantities with random components.
Matrices are written in uppercase boldface letters $\mat{C}\in\IR^{q\times{}q}$, and $\mat{C}\inv$ and $\mat{C}\T$ are its inverse and transpose, respectively. $\mat{I}_q\in\IR^{q\times q}$ is the identity square matrix. The notation $\est[\tkk-l]$ denotes an estimate at time step \(k\) conditioned on measurements up to time step \(k-l\). The Euclidean norm is denoted as \(||\vec{x}||_2= \sqrt{\vec{x}\T\vec{x}}\). The expectation and variance are given by \(\Exp\{\cdot\}\) and \(\Var\{\cdot\}\). \(|\mathbfcal{X}|\) denotes the cardinality of the set \(\mathbfcal{X}\).

\section{Event-based Triggering}
A system consisting of a sensor monitoring a linear time-invariant physical process, a triggering unit located at the sensor, and a remote receiver with a state estimator to reconstruct the physical process is considered. The setup of the event-based system is shown in Fig.~\ref{fig:system}.
Before introducing the concept of event-based transmission and estimation, the used system model will be given.
\subsection{System Model}
\label{sec:systemModel}
The state and the measurement equation of the observed process are given by the discrete-time linear system
\begin{align}
	\state[\nk] &= \sysmat[]\,\state[k] + \sysnoise[k]\,, \label{eq:sys_eq}\\
	\meas[k] &= \measmat[]\,\state[k] + \measnoise[k]   \,, \label{eq:meas_eq}
\end{align}
where \(\state[k]\in\smash{\IR^{n_x}}\) is the state at time step \(k\in\IN\), and \(\meas[k]\in\smash{\IR^{n_y}}\) denote the observations. The time-invariant process and measurement matrices are given by \(\sysmat[]\in\IR^{n_x\times{}n_x}\) and \(\measmat[]\in\IR^{n_y\times{}n_x}\), respectively. 
The process noise \(\sysnoise[l]\sim\Gauss(\zero,\syscov[])\) and measurement noise \(\measnoise[m]\sim\Gauss(\zero,\meascov[])\) are white and mutually uncorrelated for arbitrary~\(l,m\in\IN\). Further, detectability of the pair \((\sysmat,\measmat)\) is assumed. 
\subsection{Stochastic and Deterministic Event-triggering}
\label{sec:EventTrigger}
Usually, either stochastic or deterministic event triggers are considered. In this section, we provide a unified formulation for both, which allows to investigate their properties in detail.

For any innovation-based event trigger, a triggering variable \(\rvec{z}_k = \meas_k - \triggerconst[k]\) with arbitrary implicit information \(\triggerconst[k]\in\smash{\IR^{n_y}}\) needs to be defined. The triggering variable \(\rvec{z}_k\) can be understood as the error introduced by the event trigger.

Many deterministic policies are given in the form
\begin{align}
\gamma_k =
    \begin{cases}
	1, & ||\rvec{z}_k||_2 > \delta\,,\\
	0, & ||\rvec{z}_k||_2 \le \delta\,,
    \end{cases}
\end{align}
which can be rewritten to 
\begin{align}
    \gamma_k =
    \begin{cases}
	1, & \rvec{z}_k\T\triggermat\inv\rvec{z}_k > 1\,,\\
	0, & \rvec{z}_k\T\triggermat\inv\rvec{z}_k \le 1\,
    \end{cases}
    \label{eq:decisionEllipsoid}
\end{align}
with the special choice \(\triggermat=\delta^2\cdot\mat{I}\).
The transmission variable \(\gamma_k = 1\) implies an event has been triggered at time instant \(k\). More generally, the policy can be interpreted to trigger an event if \(\rvec{z}_k\) lies outside an \(n_y\)-dimensional ellipsoid with any positive definite shape matrix \(\triggermat\)~\cite{sijsEventBasedState2012}.

For stochastic triggers, a decision scheme of the form 
\begin{align}
\label{eq:decision}
    \gamma_k =
    \begin{cases}
	1, & \rv{\xi}_k > \phi(\rvec{z}_k)\,,\\
	0, & \rv{\xi}_k \le \phi(\rvec{z}_k)\,
    \end{cases}
\end{align}
is used, where \(\rv{\xi}_k\sim \mathcal{U}(0,1)\) is a random variable. 
A popular choice of the shaping function \(\phi(\rvec{z}_k):\, \IR^{n_y} \rightarrow [0,1]\) is 
\begin{align}
    \phi_\text{S}(\rvec{z}_k) =\exp\left(-\frac{1}{2}\,\rvec{z}_k\T\,\triggermat[]\inv\,\rvec{z}_k\right) \label{eq:shapingFct}
\end{align}
such that it has the form of an unnormalized Gaussian distribution, which will prove beneficial in the estimator design.
By choosing \(\phi(\rvec{z}_k) = \phi_\text{D}(\rvec{z}_k)\) with
\begin{align}
\label{eq:detTriggerPhi}
\phi_\text{D}(\rvec{z}_k) = 
\begin{cases}
    0, & \rvec{z}_k\T\triggermat\inv\rvec{z}_k > 1\,,\\
    1, & \rvec{z}_k\T\triggermat\inv\rvec{z}_k \le 1\,
\end{cases}
\end{align}
in~\eqref{eq:decision}, the deterministic triggering condition~\eqref{eq:decisionEllipsoid} can be represented in the stochastic decision scheme.

\subsection{Generalized Gaussian Weighting Function}
\label{sec:generalGaussianPhi}
To provide a single weighting function for both, deterministic and stochastic triggering policies, we use the unnormalized \ac{GND} and obtain the novel shaping function
\begin{align}
   \phi_\beta(\rvec{z}_k) = \exp{\left(-\frac{1}{2}\sqrt{\rvec{z}_k\T\triggermat\inv\rvec{z}_k}^\beta\right)}\,, 
   \label{eq:phi_beta}
\end{align}
which includes an additional parameter \(\beta>0\).
For \(\beta \rightarrow \infty\), the \ac{GND} approaches a uniform distribution as in~\eqref{eq:detTriggerPhi}, \(\beta=2\) leads to the standard Gaussian-like choice~\eqref{eq:shapingFct} of the shaping function for stochastic triggers. Moreover,~\eqref{eq:phi_beta} exposes a whole variety of shaping functions between the deterministic shaping function and the "classical" stochastic shaping function by choosing \(\beta \in [2,\infty)\). Choosing \(\beta<2\) leads to undesirable behavior due to the resulting heavy tails, as in this case small errors (values of \(\rvec{z}_k\)) trigger an event more likely and large errors less likely than with \(\beta=2\). Different choices of \(\phi_\beta\) between the extreme cases \(\beta=2\) and \(\beta\rightarrow \infty\) are shown in Fig.~\ref{fig:WeightingFct}, where 
\begin{align}
\phi(\tilde{\rvec{z}}_k) = \exp{\left(-\frac{1}{2}\sqrt{\tilde{\rvec{z}}_k\T\tilde{\rvec{z}}_k}^\beta\right)}
\end{align}
with \(\tilde{\rvec{z}}_k=\triggermat^{-\frac{1}{2}}\rvec{z}_k\) is plotted for scalar \(\rvec{z}_k\).
The positive definite design variable \(\smash{\triggermat}\in\IR^{n_y\times n_y}\) is used to control the transmission frequency, where high values of \(\triggermat\) lead to fewer transmissions.
\begin{figure}[t]
    \centering
    \setlength\fheight{\figheight}
    \setlength\fwidth{\linewidth}
    \input{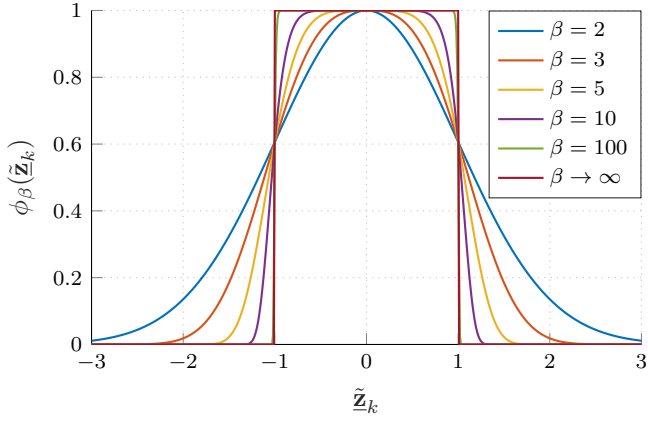}
    \caption{Univariate shaping function with varied \(\beta\).}
    \label{fig:WeightingFct}
\end{figure}
 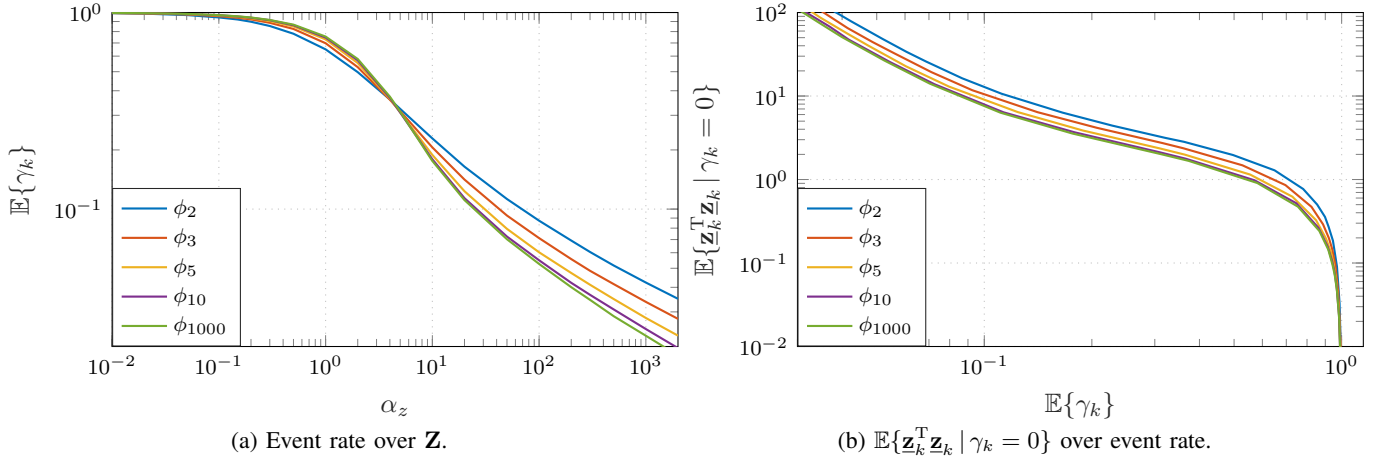
\begin{figure*}[t]
    \setlength\fheight{\figheight}
    \setlength\fwidth{\linewidth}
\begin{subfigure}{0.5\textwidth}\centering
%
%
\definecolor{mycolor1}{rgb}{0.00000,0.44700,0.74100}%
\definecolor{mycolor2}{rgb}{0.85000,0.32500,0.09800}%
\definecolor{mycolor3}{rgb}{0.92900,0.69400,0.12500}%
\definecolor{mycolor4}{rgb}{0.49400,0.18400,0.55600}%
\definecolor{mycolor6}{rgb}{0.46600,0.67400,0.18800}%
\definecolor{mycolor5}{rgb}{0.30100,0.74500,0.93300}%
\definecolor{mycolor7}{rgb}{0.63529,0.07843,0.18431}%
\begin{tikzpicture}

\begin{axis}[%
width=\textwidth,
height=\fheight,
at={(0,0)},
xmode=log,
xmin=0.01,
xmax=2000,
xminorticks=true,
xlabel style={font=\color{white!15!black}},
xlabel={$\alpha_z$\vphantom{$\Exp\{\gamma_k\}$}},
ymode=log,
ymin=0.02,
ymax=1,
yminorticks=true,
ylabel style={font=\color{white!15!black}},
ylabel={$\Exp\{\gamma_k\}$},
axis background/.style={fill=white},
legend style={at={(axis cs:1e-2, 2e-2)}, anchor=south west, legend cell align=left, align=left, draw=white!15!black}
]
\addplot [color=mycolor1]
  table[row sep=crcr]{%
0.01	0.992543999999995\\
0.02	0.987003999999994\\
0.05	0.970728\\
0.1	0.944188\\
0.15	0.921703999999997\\
0.2	0.897680000000003\\
0.3	0.854100000000002\\
0.5	0.780076000000001\\
1	0.6496\\
2	0.497328\\
4	0.361412\\
5	0.323292\\
10	0.22888\\
20	0.164368\\
50	0.11252\\
100	0.087304\\
200	0.069192\\
300	0.060512\\
500	0.0516440000000001\\
1000	0.0422999999999999\\
2000	0.034972\\
};
\addlegendentry{$\phi_2$}

\addplot [color=mycolor2]
  table[row sep=crcr]{%
0.01	0.993911999999996\\
0.02	0.989995999999994\\
0.05	0.978255999999994\\
0.1	0.958092000000005\\
0.15	0.939987999999998\\
0.2	0.922239999999998\\
0.3	0.886395999999999\\
0.5	0.824939999999999\\
1	0.697851999999999\\
2	0.527456\\
4	0.360212\\
5	0.314668\\
10	0.206016\\
20	0.141312\\
50	0.0927279999999999\\
100	0.07118\\
200	0.05568\\
300	0.048548\\
500	0.0415319999999999\\
1000	0.033728\\
2000	0.027648\\
};
\addlegendentry{$\phi_3$}

\addplot [color=mycolor3]
  table[row sep=crcr]{%
0.01	0.994659999999997\\
0.02	0.991375999999995\\
0.05	0.981079999999994\\
0.1	0.966296000000003\\
0.15	0.950360000000004\\
0.2	0.934307999999996\\
0.3	0.904656000000002\\
0.5	0.849828000000002\\
1	0.729948\\
2	0.55356\\
4	0.363808\\
5	0.310296\\
10	0.188636\\
20	0.12326\\
50	0.079356\\
100	0.060288\\
200	0.047316\\
300	0.041224\\
500	0.034872\\
1000	0.027936\\
2000	0.0227119999999999\\
};
\addlegendentry{$\phi_5$}

\addplot [color=mycolor4]
  table[row sep=crcr]{%
0.01	0.995087999999997\\
0.02	0.991963999999995\\
0.05	0.982687999999993\\
0.1	0.969408000000002\\
0.15	0.954484000000005\\
0.2	0.941031999999999\\
0.3	0.914083999999999\\
0.5	0.861472000000001\\
1	0.74584\\
2	0.570968\\
4	0.368096\\
5	0.309764\\
10	0.179308\\
20	0.11366\\
50	0.072484\\
100	0.0549079999999999\\
200	0.0421679999999999\\
300	0.0366399999999999\\
500	0.0308760000000001\\
1000	0.0245359999999999\\
2000	0.019636\\
};
\addlegendentry{$\phi_{10}$}


\addplot [color=mycolor6]
  table[row sep=crcr]{%
0.01	0.995091999999997\\
0.02	0.992575999999995\\
0.05	0.984099999999993\\
0.1	0.969891999999999\\
0.15	0.956508000000004\\
0.2	0.943564\\
0.3	0.916379999999999\\
0.5	0.867623999999999\\
1	0.754651999999999\\
2	0.58024\\
4	0.373316\\
5	0.313\\
10	0.175736\\
20	0.111112\\
50	0.0702559999999999\\
100	0.052604\\
200	0.0400679999999999\\
300	0.03454\\
500	0.0285200000000001\\
1000	0.0226879999999999\\
2000	0.018016\\
};
\addlegendentry{$\phi_{1000}$}


\end{axis}

\end{tikzpicture}%
  \caption{Event rate over $\triggermat$.\vphantom{$\Exp\{\rvec{z}\T\}$}}
 \label{fig:EventRateZConstVel}
 \end{subfigure}
 \begin{subfigure}{0.5\textwidth}\centering
%
%
\definecolor{mycolor1}{rgb}{0.00000,0.44700,0.74100}%
\definecolor{mycolor2}{rgb}{0.85000,0.32500,0.09800}%
\definecolor{mycolor3}{rgb}{0.92900,0.69400,0.12500}%
\definecolor{mycolor4}{rgb}{0.49400,0.18400,0.55600}%
\definecolor{mycolor5}{rgb}{0.46600,0.67400,0.18800}%
\definecolor{mycolor6}{rgb}{0.30100,0.74500,0.93300}%
\begin{tikzpicture}

\begin{axis}[%
width=\textwidth,
height=\fheight,
at={(0,0)},
xmode=log,
xmin=0.03,
xmax=1.15,
xminorticks=true,
xlabel style={font=\color{white!15!black}},
xlabel={$\Exp\{\gamma_k\}$},
ymode=log,
ymin=0.01,
ymax=100,
yminorticks=true,
ylabel style={font=\color{white!15!black}},
ylabel={$\Exp\{\rvec{z}_k\T\rvec{z}_k\,|\,\gamma_k=0\}$},
axis background/.style={fill=white},
legend style={at={(axis cs:3e-2, 1e-2)}, anchor=south west, legend cell align=left, align=left, draw=white!15!black}
]

\addplot [color=mycolor1]
  table[row sep=crcr]{%
0.992163999999995	0.01320615899257\\
0.986823999999994	0.032607914718898\\
0.970304	0.0905174762450541\\
0.945392000000001	0.185265794877054\\
0.919879999999997	0.269749568857095\\
0.898240000000002	0.356837016050103\\
0.854036000000001	0.501691052175465\\
0.780608000000001	0.770170738520044\\
0.64902	1.2887992965608\\
0.496408	1.96657094056383\\
0.362392	2.82638212255128\\
0.320816	3.15381663327991\\
0.228044	4.42168259317512\\
0.165716	6.31946399556433\\
0.111756	10.653898356015\\
0.0867359999999999	16.2938162497088\\
0.0687400000000001	25.8778733247172\\
0.0605360000000001	33.870277792756\\
0.05192	48.2831045572897\\
0.0423119999999999	78.8491408016638\\
0.034756	129.245344525414\\
0.0229719999999999	407.560638518406\\
0.0153680000000001	1331.04887417752\\
0.00758800000000003	11696.4350639459\\
};
\addlegendentry{$\phi_{2}$}

\addplot [color=mycolor2, ]
  table[row sep=crcr]{%
0.994003999999996	0.0059530543216919\\
0.989931999999994	0.0159773418153312\\
0.977915999999994	0.0458174823262295\\
0.959532000000005	0.0969035949850539\\
0.940531999999998	0.147650357902921\\
0.921399999999997	0.198045733545507\\
0.88698	0.290883370391525\\
0.823252	0.471304570167845\\
0.696256	0.857127850432518\\
0.527856	1.47909300417159\\
0.360468	2.36864582645978\\
0.31544	2.73050295456936\\
0.20648	4.1640526316264\\
0.140928	6.40553609395511\\
0.0925999999999999	11.7250673770353\\
0.071168	19.3016493133043\\
0.055832	32.6853635296076\\
0.048712	44.2392206612026\\
0.041444	65.1062603508\\
0.033908	113.681939058683\\
0.027572	193.631730842903\\
0.017464	695.151427195187\\
0.011636	2519.10053116337\\
0.005728	26506.6813991062\\
};
\addlegendentry{$\phi_{3}$}

\addplot [color=mycolor3]
  table[row sep=crcr]{%
0.994591999999997	0.00347163646461676\\
0.991487999999995	0.00979789136539184\\
0.981799999999993	0.0297353204222649\\
0.964972000000004	0.064286932534333\\
0.949640000000003	0.0980165199254237\\
0.934635999999996	0.132882280098175\\
0.904580000000003	0.195545921533792\\
0.847876000000002	0.326164150329361\\
0.728112	0.623004222933207\\
0.554464	1.14657316027587\\
0.364344	1.98445661964033\\
0.31114	2.34734289813887\\
0.188704	3.86473408655545\\
0.12406	6.42150302117605\\
0.079612	12.9856946623787\\
0.060352	22.7069588531377\\
0.0473199999999999	40.2144449169386\\
0.041168	56.6748940331963\\
0.03474	87.1117734683369\\
0.0282200000000001	157.851558340455\\
0.0225439999999999	285.929058595978\\
0.014156	1125.6975291996\\
0.00924800000000001	4549.80442253258\\
0.00442800000000002	55048.0616509113\\
};
\addlegendentry{$\phi_{5}$}

\addplot [color=mycolor4]
  table[row sep=crcr]{%
0.995087999999997	0.00264513207290092\\
0.992115999999995	0.00723573373249712\\
0.983095999999993	0.0236061136871186\\
0.968564000000002	0.0511277626890449\\
0.954620000000005	0.0791591921026905\\
0.940391999999999	0.106095908055714\\
0.912972	0.159013471850168\\
0.860624	0.264994223432931\\
0.747127999999999	0.518971729492964\\
0.570651999999999	0.978299826326538\\
0.366444	1.77377564311834\\
0.30954	2.11951302908818\\
0.179152	3.67150245570323\\
0.113256	6.37552539710765\\
0.0725800000000001	13.8318646009339\\
0.0545919999999999	25.3732629606564\\
0.0417559999999999	47.3386371380705\\
0.036888	67.9384905947859\\
0.03072	108.768578373637\\
0.0243639999999999	203.244667309307\\
0.019572	389.19553928376\\
0.01208	1695.70864639774\\
0.00761200000000004	7242.64915538315\\
0.00362800000000003	99123.0617169327\\
};
\addlegendentry{$\phi_{10}$}


\addplot [color=mycolor5]
  table[row sep=crcr]{%
0.995299999999997	0.00218812264159826\\
0.992311999999995	0.00664469432877888\\
0.984171999999993	0.0216506831934307\\
0.970483999999999	0.0463464180194509\\
0.956824000000005	0.0713654370327867\\
0.943075999999999	0.0949185055655454\\
0.917699999999997	0.144245820420151\\
0.866419999999999	0.241653024557547\\
0.754611999999999	0.472914793190811\\
0.580748	0.908337710602555\\
0.37334	1.67171920484704\\
0.312952	2.00803580201577\\
0.176484	3.55517419496474\\
0.110968	6.31775708492549\\
0.070368	14.0838213116903\\
0.0524679999999999	26.6747085660173\\
0.039864	51.1595011858204\\
0.034456	75.6464689284929\\
0.0286920000000001	124.384103473246\\
0.0225759999999999	243.98453326193\\
0.017928	468.439294055844\\
0.010736	2258.05227403582\\
0.00677600000000002	10479.1948398871\\
0.00329600000000003	165907.315477272\\
};
\addlegendentry{$\phi_{1000}$}


\end{axis}

\end{tikzpicture}%
  \caption{$\Exp\{\rvec{z}_k\T\rvec{z}_k\,|\,\gamma_k=0\}$ over event rate.}
 \label{fig:MeanZk}
 \end{subfigure}
 \caption{Trigger performance with 2D constant velocity system model.}
 \label{fig:EventRateZ}
 \end{figure*}

The transmission probability conditioned on \(\meas_k\) is given by 
\begin{align}
	p_k^\text{Tx} = \Pr\bigl\{\gamma_k = 1\,\big|\,\meas_k\bigr\} &= 1 - \phi_\beta(\meas_k - \triggerconst[k])\,, \label{eq:likelihood_triggered}\\
	 1-p_k^\text{Tx} = \Pr\bigl\{\gamma_k = 0\,\big|\,\meas_k\bigr\} &= \phi_\beta(\meas_k - \triggerconst[k])\,. \label{eq:likelihood_not_triggered}
\end{align}

\subsection{Popular Choices of \(\triggerconst_k\)}
\label{sec:choiceCk}
Generally, the choice of the implicit measurement \(\triggerconst[k]\) is arbitrary, it could even be set to 0. However, its evolution \(\triggerconst_k = g(\triggerconst[k-1])\) needs to be known to the sensor and the remote estimator. Depending on the observed system, the \ac{SOD} and the \ac{SODP} scheme have proven to efficiently reduce the data rate while maintaining acceptable estimation performance. 
\subsubsection{Send-on-Delta}
In the \ac{SOD} scheme, \(\triggerconst[k]\) is set to the measurement \(\meas[k_e]\) transmitted in the last event instant \(k_e\). Hence, \(\rvec{z}_k\) is given by
\begin{align}
    \rvec{z}_k = \meas_k - \meas_{k_e}\,.
\end{align}
An obvious advantage of this scheme is its simplicity; however, in unstable systems, the transmission rate will vary over time.

\subsubsection{Send-on-Delta with Prediction}
In the \ac{SODP} scheme, \(\triggerconst[k]\) is set to a prediction of the local state estimate of the sensor \(\est[k_e]^\text{S}\) transmitted in the last event instant \(k_e = k-l\), which was \(l\) time steps ago. Accordingly, \(\rvec{z}_k\) is given by
\begin{align}
    \rvec{z}_k = \meas_k - \measmat\sysmat^l\est^\text{S}_{k_e}\,.
\end{align}
This scheme requires a local state estimator at the sensor to obtain \(\est^\text{S}_{k_e}\) as depicted in Fig.~\ref{fig:system}. Fortunately, the estimation quality usually does not affect the quality of the estimation at the remote estimator, only the transmission rate.

Further details and advantages and disadvantages of these choices are discussed in~\cite{hanStochasticEventTriggeredSensor2015,andrenEventBasedStateEstimation2016, schmittGaussianityPreservingEventBasedState2019}.

\subsection{Choice of \(\beta\)}
\label{sec:ChoicePhi}
With the introduction of \(\beta\) in \(\phi_\beta(\rvec{z}_k)\), a new degree of freedom has been introduced in the trigger design. This opens the question of how \(\beta\) should be chosen. 
In this section, the question is approached from the trigger side without considering possible state reconstruction at a remote estimator.

Some properties of the different weighting functions can be directly deduced from Fig.~\ref{fig:WeightingFct}. The weighting functions for different values of \(\beta\) all intersect in two points, \(\tilde{\rvec{z}}_k=1\) and \(\tilde{\rvec{z}}_k=-1\).
For \(|\tilde{\rvec{z}}_k| < 1\) the weighting functions with \(\beta < \infty\) lead to a higher transmission rate than \(\beta = \infty\), for \(|\tilde{\rvec{z}}_k| > 1\) the weighting functions with \(\beta < \infty\) lead to a lower transmission rate than \(\beta = \infty\), as the transmission probability per time step is given by~\eqref{eq:likelihood_triggered}. 
To further explore the properties of different choices of \(\beta\), Monte Carlo simulations with 500 runs and 500 time steps each are performed for a 2D linear nearly constant velocity system model with the choice \(\triggermat = \alpha_z\cdot\mat{I}\) and \ac{SODP} for the choice of \(\triggerconst_k\) (cf. Section~\ref{sec:Simulation}). Ergodicity is assumed for the approximation of the expectations~\cite{hanStochasticEventTriggeredSensor2015}. 
In Fig.~\ref{fig:EventRateZConstVel}, the event rate is plotted over \(\alpha_z\) to show the development of the transmission rates discussed before. But lowering the mean transmission rate is not the only objective of event-based triggering, the trade-off between transmission rate and estimation error at the remote estimator is crucial. Hence, the error introduced by the triggering policy is considered to obtain an indication on the achievable estimation quality. The error is given by \(\rvec{z}_k = \meas_k - \triggerconst[k]\). Since the weighting function uses only the square of \(\rvec{z}_k\), \(\rvec{z}_k\T\triggermat\inv\rvec{z}_k\), and \(\triggermat\) is positive definite, it is sufficient to consider \(\Exp\{\rvec{z}_k\T\rvec{z}_k\}\). In Fig.~\ref{fig:MeanZk}, \(\Exp\{\rvec{z}_k\T\rvec{z}_k\}\) is plotted over the transmission rate. It can be observed that, as expected, the squared error decreases monotonically for increasing transmission rates and high values of \(\beta\) always lead to lower errors than lower values of \(\beta\).\looseness=-1

From the above considerations it can be concluded that deterministic triggers generally have a better performance than stochastic triggers. This result is also in line with the investigations of~\cite{yuStochasticDeterministicEventBased2021}, who proved the existence of a better deterministic trigger for any stochastic trigger in the scalar linear case.
However, the benefits of the deterministic triggering policies can only be exploited if suitable consistent estimators are available that incorporate the implicit information in non-transmission instants efficiently.

Before investigating the estimator design, the event-based procedure is reviewed.

\subsection{Event-based Procedure}
The event-based procedure with a stochastic trigger can be summarized as follows.
\subsubsection{\(\gamma_k = 1\)}
An event is triggered at the sensor in time step \(k=k_e\). Thus, the current measurement \(\meas[k_e]\) and, if \(\triggerconst[k_e]\neq\meas[k_e]\), \(\triggerconst[k_e]\)
are transmitted to the receiving remote estimator. The remote estimator performs a filtering step with the received measurement. 

\subsubsection{\(\gamma_k = 0\)}
Both, sensor and receiver use the same rule to determine \(\triggerconst[k]\) from \(\triggerconst[k_e]\) of the last event instant. 
The estimator uses the implicit measurement \(\tilde{\meas}_k = \triggerconst[k]\) for its update step.

Suitable estimators will be explored in the following.

\section{Event-based Estimation}
Two different estimators are presented, the \ac{SEBKF}~\cite{hanStochasticEventTriggeredSensor2015,andrenEventBasedStateEstimation2016, schmittGaussianityPreservingEventBasedState2019} specifically designed for stochastic triggers with Gaussian weighting functions \(\phi_2({\rvec{z}}_k)\) and a sampling-based approach that is by design not restricted to any kind of weighting function.

\subsection{Stochastic Event-based Kalman Filter}
\label{sec:MatchingEstimator}
To exploit the implicit information conveyed in non-transmission instants, suitable estimators are required. The implicit measurement \(\tilde{\meas}_k = \measmat[]\,\state[k] + \measnoise[k] + \rvec{\nu}_k\) shall be used in non-transmission instants (\(\gamma_k=0\)), where \(\triggerconst[k]\) is considered as realization of \(\tilde{\meas}_k\). Here, \(\rvec{\nu}_k\) denotes the additional noise introduced by the triggering condition.
For \(\beta=2\) and linear \acp{SSM}, such an estimator is available, and is a direct extension of a standard linear \ac{KF} for arbitrary choices of \(\triggerconst[k]\) as shown in~\cite{schmittGaussianityPreservingEventBasedState2019}.
The resulting periodic estimator, \ac{SEBKF}, is given in the following. For the prediction of the state estimate \(\est[\tkk]\) and estimation error covariance \(\estcov[\tkk]\), the standard prediction step
\begin{subequations} \label{eq:KF_pred}
\begin{align}
\est[\tkpk] &=  \sysmat[]\,\est[k-1|k-1]  \, ,  \label{eq:KF_pred_est} 
\\
\estcov[\tkpk] &=  \sysmat[]\,\estcov[k-1|k-1]\,\sysmat[]\T + \syscov[]  \,  \label{eq:KF_pred_cov}
\end{align}
\noeqref{eq:KF_pred_est}
\noeqref{eq:KF_pred_cov}
\end{subequations}
is used. In the update step given by
\begin{subequations}\label{eq:KF_update}
\begin{align}
\est[\tkk] &= \est[\tkpk]+ \gain[k]\,(\gamma_k\, \rvec{z}_k-\hat{\rvec{z}}_{k|k-1})\,,\label{eq:KF_up_est}\\
\rvec{z}_k &= \rvec{y}_k - \triggerconst[k]\,,\quad
\hat{\rvec{z}}_{k|k-1} = \measmat[]\,\est[\tkpk] - \triggerconst[k]\,, \label{eq:triggerVariable}\\
\estcov[\tkk] &=  \bigl(\mat{I}_{n_x}-\gain[k]\,\measmat[]\bigr)\,\estcov[\tkpk]  \, , \label{eq:KF_up_cov}
\intertext{with the Kalman gain}
\gain[k] &= \estcov[\tkpk]\,\measmat[]\T\Bigl(\measmat[]\estcov[\tkpk]\,\measmat[]\T + \meascov + (1-\gamma_k)\triggermat[]\Bigr)\inv \hspace{5pt} \label{eq:KF_up_gain}
\end{align}
\end{subequations}
the innovation term and the measurement noise covariance are altered.
\noeqref{eq:KF_up_est}
\noeqref{eq:triggerVariable}
\noeqref{eq:KF_up_cov}
\noeqref{eq:KF_up_gain}
For \(\gamma_k = 1\), in a transmission instant, \(\triggerconst[k]\) and \(\triggermat\) vanish and the above equations reduce to those of the standard \ac{KF}. If no event is triggered and hence \(\gamma_k = 0\), \(\triggerconst[k]\) is considered as an implicit measurement.
The estimator yields the optimal estimation result under the provided information if \(\beta=2\). 

\begin{figure}[b]
    \centering
    \setlength\fheight{\figheight}
    \setlength\fwidth{\linewidth}
%
%
\definecolor{mycolor1}{rgb}{0.00000,0.44700,0.74100}%
\definecolor{mycolor2}{rgb}{0.85000,0.32500,0.09800}%
\definecolor{mycolor3}{rgb}{0.92900,0.69400,0.12500}%
\definecolor{mycolor4}{rgb}{0.49400,0.18400,0.55600}%
\definecolor{mycolor5}{rgb}{0.46600,0.67400,0.18800}%
\definecolor{mycolor6}{rgb}{0.30100,0.74500,0.93300}%
\definecolor{mycolor7}{rgb}{0.63529,0.07843,0.18431}%
\begin{tikzpicture}

\begin{axis}[%
width=\fwidth,
height=\fheight,
at={(0,0)},
xmode=log,
xmin=0.01,
xmax=5e5,
xminorticks=true,
xlabel style={font=\color{white!15!black}},
xlabel={$\alpha_z$},
ymode=log,
ymin=1e-3,
ymax=1e5,
yminorticks=true,
ylabel style={font=\color{white!15!black}},
ylabel={\(\Exp\{\rvec{z}_k\T\rvec{z}_k\,|\,\gamma_k=0\}\)},
axis background/.style={fill=white},
legend style={at={(axis cs:5e5, 1e-3)}, anchor=south east, legend cell align=left, align=left, draw=white!15!black}
]
\addplot [color=mycolor1]
  table[row sep=crcr]{%
0.01	0.0136162550532072\\
0.02	0.0324483507758898\\
0.05	0.0835534059413218\\
0.1	0.186951639154245\\
0.15	0.270247448354835\\
0.2	0.349518650522241\\
0.3	0.506972704482062\\
0.5	0.765622863727449\\
1	1.29043633536019\\
2	1.98626022036281\\
4	2.79515262482468\\
5	3.16606173716951\\
10	4.40499525522205\\
20	6.44829656398942\\
50	10.702388790444\\
100	16.2582995510168\\
200	26.2941528348196\\
300	33.5561605048646\\
500	49.2510887368914\\
1000	78.8246871555678\\
2000	128.021363358027\\
10000	405.624008380416\\
50000	1397.58560946529\\
1000000	11892.8629747877\\
};
\addlegendentry{$\phi_2$}

\addplot [color=mycolor2]
  table[row sep=crcr]{%
0.01	0.00554968104629071\\
0.02	0.0150485337499302\\
0.05	0.0471630867191534\\
0.1	0.0956092018371417\\
0.15	0.149138246471624\\
0.2	0.195132541608997\\
0.3	0.29622541602031\\
0.5	0.471241177860662\\
1	0.878660109491849\\
2	1.47654635783707\\
4	2.34448862783343\\
5	2.72787680895271\\
10	4.12271516283328\\
20	6.42091693319603\\
50	11.8171342701678\\
100	19.4704515221527\\
200	32.3903912870201\\
300	44.9137110383832\\
500	64.5638450461295\\
1000	111.72124563006\\
2000	194.105157817313\\
10000	692.545317340209\\
50000	2494.75506378911\\
1000000	26288.6576056891\\
};
\addlegendentry{$\phi_3$}

\addplot [color=mycolor3]
  table[row sep=crcr]{%
0.01	0.00332203027451843\\
0.02	0.00951925080091758\\
0.05	0.0318763571356845\\
0.1	0.0638630716454265\\
0.15	0.0972490856866442\\
0.2	0.132812909990417\\
0.3	0.194594300235577\\
0.5	0.335641367079514\\
1	0.621387619274869\\
2	1.1543458647959\\
4	1.98289936300354\\
5	2.35476842725578\\
10	3.84912411429139\\
20	6.3713999182694\\
50	13.1626816879529\\
100	22.6092848351316\\
200	39.3303535377664\\
300	56.1157855805852\\
500	85.2738981361906\\
1000	158.308149344639\\
2000	277.512277122204\\
10000	1163.36073121068\\
50000	4398.26579268401\\
1000000	51551.6659107133\\
};
\addlegendentry{$\phi_5$}

\addplot [color=mycolor4]
  table[row sep=crcr]{%
0.01	0.00273032760295457\\
0.02	0.00762927348674072\\
0.05	0.0223447208763658\\
0.1	0.0533201084709865\\
0.15	0.0778282227924659\\
0.2	0.105100084949431\\
0.3	0.16106986443611\\
0.5	0.267550131386584\\
1	0.513098105835041\\
2	0.985397058050424\\
4	1.75535276093204\\
5	2.1231251163987\\
10	3.66131359323036\\
20	6.44515644799537\\
50	13.9384245966173\\
100	25.8806300726973\\
200	47.0007954823935\\
300	67.8225061298564\\
500	110.486470575535\\
1000	206.064331672958\\
2000	381.5753753794\\
10000	1641.39428892551\\
50000	7114.35587804025\\
1000000	102905.392700468\\
};
\addlegendentry{$\phi_{10}$}


\addplot [color=mycolor5]
  table[row sep=crcr]{%
0.01	0.00209800406184815\\
0.02	0.00622668962742753\\
0.05	0.0230920971545255\\
0.1	0.0465649110086126\\
0.15	0.0718680764525079\\
0.2	0.0961965166270902\\
0.3	0.143198571801646\\
0.5	0.237787366965978\\
1	0.470709419927163\\
2	0.891014490820901\\
4	1.66462917193563\\
5	2.01165784705506\\
10	3.54495122218918\\
20	6.28875716927018\\
50	13.9433706861829\\
100	26.6147442252687\\
200	51.1448404264332\\
300	76.2874478650526\\
500	122.28178094288\\
1000	247.724607355266\\
2000	470.945401477698\\
10000	2366.32240345979\\
50000	10177.1754576859\\
1000000	158852.997125153\\
};
\addlegendentry{$\phi_{1000}$}


\end{axis}
\end{tikzpicture}%
    \caption{\(\Exp\{\rvec{z}_k\T\rvec{z}_k\,|\,\gamma_k=0\}\) over $\triggermat$ for a 2D nearly constant velocity system model.}
    \label{fig:zZz_ZConstVel}
 \end{figure}

However, for other choices of \(\beta\),
\ac{SEBKF} will generally not be optimal and might even lead to inconsistent estimation results, where the estimated error covariance is lower than the actual covariance of the error.
To find consistent estimators for generalized Gaussian weighting functions with \(\beta>2\), one could try to bound the error \(\rv{\nu}_k\) with the help of a Gaussian distribution and use the estimator~\eqref{eq:KF_pred} --~\eqref{eq:KF_update}. Unfortunately, this is not possible, because the noise process \(\{\rvec{\nu}_{k}\}_{k\in K}\,, K=k_e,\dots,k_{e+1}\) is generally correlated in time in inter-event sequences since \(\triggerconst[k]\) depends on former time steps. Hence, the problem aggravates for long inter-event sequences, i.\,e., large values in \(\triggermat\). 

Nevertheless, according to simulation studies, \ac{SEBKF} is a conservative estimator for small enough choices of \(\triggermat\) such that \(\Exp\{\rvec{z}_k\T\rvec{z}_k\,|\,\gamma_k=0, \beta>2\}\leq \Exp\{\rvec{z}_k\T\rvec{z}_k\,|\,\gamma_k=0, \beta=2\}\) holds. This is the case because the error introduced by triggering policies using \(\beta>2\) is then on average lower than with \(\beta=2\) and hence \(\meascov + \triggermat\) conservatively bounds the measurement error \(\measnoise_k+\rv{\nu}_k\) of the implicit measurement.
The maximum value \(\triggermat^*\) for which this holds can be determined graphically evaluating Fig.~\ref{fig:zZz_ZConstVel}, which was obtained using the same system simulation as in Section~\ref{sec:ChoicePhi}.
Considering Fig.~\ref{fig:zZz_ZConstVel} and Fig.~\ref{fig:EventRateZConstVel}, it can be concluded that the \ac{SEBKF} estimator will be consistent up to \(\triggermat^*\approx20\cdot \mat{I}\) or a mean transmission rate of \(\bar{\gamma}\approx0.1\), which will be confirmed in the system simulations in section~\ref{sec:Simulation}.

To achieve consistent estimates at low transmission rates, a more versatile estimator will be presented in the following.

\subsection{Sampling-based Estimator}
\label{sec:SamplingEstimator}
To overcome the limitations of the \ac{SEBKF} presented in section~\ref{sec:MatchingEstimator}, an event-based sampling-based estimator building on the idea of \acl{ABC} and particle filters~\cite{turnerTutorialApproximateBayesian2012, siggesLikelihoodFreeParticleFilter2017} is introduced. The presented estimator for stochastic triggers is similar to the one proposed by~\cite{gasmiEventTriggeredStateEstimation2022} for deterministic triggers with some improvements.
The developed estimator is based on the following considerations:
The desired posterior conditional probability distribution is \(p(\state_k\,|\,\mathcal{I}_k)\) with \(\mathcal{I}_k = \{\gamma_0,\dots,\gamma_k,\gamma_0\meas_0,\dots,\gamma_k\meas_k\}\).
In case of a triggered event, \(\gamma_k=1\), it is simply given by
\begin{align}
     &p(\state_k\,|\,\mathcal{I}_k) = p(\state_k\,|\,\meas_k,\, \mathcal{I}_{k-1})
     = \frac{ p(\meas_k\,|\,\state_k)\, p(\state_k\,|\,\mathcal{I}_{k-1})}{p(\meas_k\,|\,\mathcal{I}_{k-1})} \\
     &\propto \int\limits_{-\infty}^\infty p(\meas_k\,|\,\state_k)\, p(\state_k\,|\,\state_{k-1})\, p(\state_{k-1}\,|\,\mathcal{I}_{k-1})\operatorname{d}\state[k-1]\,,
\end{align}
which leads to the standard linear \ac{KF} for linear \acp{SSM}. If \(\gamma_k=0\), the triggering policy has to be taken into account to exploit the implicit information. The update step can be written as
\begin{align}
     &p(\state_k\,|\,\mathcal{I}_k) = p(\state_k\,|\,\gamma_k=0,\, \mathcal{I}_{k-1}) \\
     &= \frac{p(\gamma_k=0\,|\, \state[k])\,p(\state_k\,|\,\mathcal{I}_{k-1})}{p(\gamma_k=0\,|\,\mathcal{I}_{k-1})}\\
     &= \int\limits_{-\infty}^\infty\frac{p(\gamma_k=0\,|\,\meas_k)\, p(\meas_k\,|\,\state_k)\, p(\state_k\,|\,\mathcal{I}_{k-1})}{p(\gamma_k=0\,|\,\mathcal{I}_{k-1})}\operatorname{d}\meas[k]\\
     &\propto \int\limits_{-\infty}^\infty\phi_\beta(\meas_k-\triggerconst_k)\, p(\meas_k\,|\,\state_k)\,p(\state_k\,|\,\mathcal{I}_{k-1}) \operatorname{d}\meas[k]\label{eq:distribution_gamma0}
\end{align}
considering~\eqref{eq:likelihood_not_triggered}, i.\,e.,
\begin{align}
    p(\gamma_k=0\,|\,\meas_k) = \phi_\beta(\meas_k-\triggerconst_k)\,.
\end{align}
Even for linear \acp{SSM},~\eqref{eq:distribution_gamma0} cannot be solved easily if \(\beta\neq2\). Therefore, a sampling-based approach is employed that obtains the result of~\eqref{eq:distribution_gamma0} by simulation using a set of samples and knowledge about the state and measurement equations as well as the triggering condition at the remote estimator. The procedure is described in the following and summarized in Algorithm~\ref{alg:SampleSEBKF}.

The estimator is initialized by drawing \(N\) samples \(\mathbfcal{X}^i_0\) from the initial state estimate's distribution \(\est_0 \sim \mathcal{N}(\state_0,\estcov_0)\). In the following, event instants and non-event instants are distinguished.
\subsubsection{\(\gamma_k=1\)}
If an event was triggered and \(\meas[k]\) is received in time step \(k\), a standard linear \ac{KF} prediction and update step (cf. equations~\eqref{eq:KF_pred} to~\eqref{eq:KF_update} for \(\gamma_k=1\)) are performed using \(\est[k-1|k-1]\) and \(\estcov[k-1|k-1]\).

\subsubsection{\(\gamma_k=0\)}
If no event was triggered, the implicit measurement \(\triggerconst_{k-1}\) is propagated according to \(\triggerconst_k = g(\triggerconst_{k-1})\) using the predefined propagation rule \(g(\triggerconst[k])\) (c.\,f., Section~\ref{sec:choiceCk}).
Then, if no samples are available from the last time step (i.\,e., at \(k-1\) an event was triggered), new samples \(\mathbfcal{X}^i_{k-1}\) are generated from \(\est[k-1] \sim \mathcal{N}(\state[k-1],\estcov[k-1])\).
The samples are predicted using the system equation~\eqref{eq:sys_eq} and drawing \(N\) process noise samples from \(\sysnoise[k-1] \sim \mathcal{N}(0,\syscov)\) (Algorithm~\ref{alg:SampleSEBKF} line~\ref{alg_l:sys_eq}).  
Subsequently, the measurement equation~\eqref{eq:meas_eq} is applied in a similar fashion drawing noise samples from \(\measnoise[k-1] \sim \mathcal{N}(0,\meascov)\) (line~\ref{alg_l:meas_eq}).
Then, \(\rvec{z}_k^i = \meas_k^i - \triggerconst[k]\) is calculated for each sample \(i=1,\dots,N\). Using the triggering condition with the weighting function \(\phi(\rvec{z}^i_k)\) and \(\xi_k^i\sim \mathcal{U}(0,1)\) (line~\ref{alg_l:checktriggercond}), those samples that lead to \(\gamma^i_k=1\), indicating that an event would have been triggered, are rejected (line~\ref{alg_l:reject}).
The sampling and rejection process is repeated until \(\tilde{N} \ge N\) accepted samples have been obtained. Of those samples, \(N\) are chosen randomly. If \(\tilde{N} > N\), the selection process is repeated \(M\) times and the selection with the highest variance is kept to robustify the algorithm. The accepted samples represent the new state estimate \(\est[\tkk]\).
After completing the update step, the state estimate and the corresponding estimation error covariance are obtained using
\begin{align}
    \est[\tkk] &= \frac{1}{{N}}\sum\limits_{i=1}^{{N}} \mathbfcal{X}_k^i\,,\\
    \estcov[\tkk] &=\frac{1}{{N-1}} \sum\limits_{i=1}^{{N}} \left(\mathbfcal{X}_k^i-\est[\tkk])(\mathbfcal{X}_k^i-\est[\tkk]\right)\T\,.
\end{align}
\begin{algorithm}
    \caption{Sampling-based Estimator} \label{alg:SampleSEBKF}
    \begin{algorithmic}[1]
        \REQUIRE \(\est_{k-1|k-1},\, \estcov_{k-1|k-1},\, \gamma_k\meas_k\)
        \ENSURE \(\est[\tkk], \estcov[\tkk]\)
        \IF[Event triggered]{\( \gamma_k = 1\)}
            \STATE \((\est[\tkpk],\, \estcov[\tkpk]) = \operatorname{prediction}(\est_{k-1|k-1},\, \estcov_{k-1|k-1})\)
            \STATE \(\triggerconst_k = \meas_k\)
            \STATE \((\est[\tkk],\, \estcov[\tkk]) = \operatorname{update}(\est[\tkpk],\, \estcov[\tkpk],\, \meas[k])\)      
        \ELSIF[No event triggered]{\(\gamma_k = 0\)}
            \STATE \(\triggerconst[k] = g(\triggerconst_{k-1})\)
            \IF{\(\mathbfcal{X}_{k-1}\) is not available}
                \STATE \(\mathbfcal{X}_{k-1} = \operatorname{sample}(\est_{k-1|k-1})\)
            \ENDIF
            \WHILE{\(|{\mathbfcal{X}_{k}}|< N\)}
                \FOR{\(i = 1:N\)} 
                    \STATE \(\mathbfcal{X}^i_k = \sysmat \mathbfcal{X}^i_{k-1} + \sysnoise^i_{k-1}\) \label{alg_l:sys_eq}
                    \STATE \(\meas^i_k = \measmat \mathbfcal{X}^i_{k} + \measnoise^i_{k-1}\) \label{alg_l:meas_eq}
                    \STATE \(\rvec{z}^i_k = \meas^i_k - \triggerconst[k]\)
                    \IF{\(\xi^i_k > \phi(\rvec{z}_k^i)\)}\label{alg_l:checktriggercond}
                        \STATE \(\operatorname{Reject }\mathbfcal{X}^i_k\)\label{alg_l:reject}
                    \ENDIF                   
                \ENDFOR
            \ENDWHILE
            \STATE Choose \(N\) samples from \(\mathbfcal{X}_k\) and set as \(\tilde{\mathbfcal{X}}_{k,0}\)
            \FOR {\(j = 1:M\)}
            \STATE Choose \(N\) samples from \(\mathbfcal{X}_k\) and set as \(\tilde{\mathbfcal{X}}_{k,j}\)
            \IF{\(\operatorname{Var}\{\tilde{\mathbfcal{X}}_{k,j}\}<\operatorname{Var}\{\tilde{\mathbfcal{X}}_{k,j-1}\}\)}
            \STATE \(\tilde{\mathbfcal{X}}_{k,j}\) = \(\tilde{\mathbfcal{X}}_{k,j-1}\)
            \ENDIF
            \ENDFOR
            \STATE \(\mathbfcal{X}_{k}\) = \(\tilde{\mathbfcal{X}}_{k,M}\)
        \ENDIF
        \STATE \(\est[\tkk] = \frac{1}{{N}}\sum\limits_{i=1}^{{N}} \mathbfcal{X}_k^i\)
        \STATE\(\estcov[\tkk] =\frac{1}{{N-1}} \sum\limits_{i=1}^{{N}} (\mathbfcal{X}_k^i-\est[\tkk])(\mathbfcal{X}_k^i-\est[\tkk])\T\)
    \end{algorithmic}
\end{algorithm}
\begin{remark}
    Even though the sampling and rejection procedure is repeated until sufficient samples have been obtained, this will usually not lead to excessive run times, because the fact that no event has been triggered means that the current measurement \(\meas[k]\) is sufficiently close to the current estimate \(\est[\tkpk]\) such that drawing samples that fulfill the non-triggering condition are found with high probability. Only very high transmission rates, where the non-transmission region is very tight and \(\estcov[\tkpk]\) is small, may require large numbers of samples to be drawn.
\end{remark}

\begin{remark}
\label{re:approximation}
    The procedure in event instants could be replaced by a sampling-based approach as well to avoid any Gaussian approximations in the filter. However, since the optimal prediction and update equations are known for linear systems, using them leads to better estimation results and reduces the computational burden. In this process, only one Gaussian approximation is made, namely, to generate new samples after the \ac{KF} update step, as in the \ac{KF} itself no assumptions regarding Gaussian distributions of the variables~\cite{uhlmannGaussianityKalmanFilter2022} are made. The approximation after the update step of the \ac{KF} is not very severe, because the Gaussian distribution of the measurement \(\meas[k]\) dominates the update (i.\,e., the Kalman gain will be large, cf.~\eqref{eq:KF_up_gain}), especially for low transmission rates.
    
    Nevertheless, the use of the sampling approach in event instants is beneficial if nonlinear system and/or measurement equations are considered and optimal filter equations are unavailable.
\end{remark}

The advantage of the sampling-based estimator over \ac{SEBKF} lies in the absence of approximations in non-transmission instants. The sampling-based estimator can accurately sample the area of the weighting function \(\phi_\beta(\rvec{z}_k)\) that leads to \(\gamma_k=0\) and propagate those samples in the next step if still no event was triggered. Due to the simulation-based approach, the estimator will also be directly applicable to other triggering schemes than the innovation-based Gaussian stochastic scheme described in Sections~\ref{sec:EventTrigger} and~\ref{sec:generalGaussianPhi}.

The advantages and disadvantages of both estimators will be illustrated with the help of simulation examples in the next section.\looseness=-1

\section{Simulation Results}
\label{sec:Simulation}
To evaluate the remote estimation performance of the complete event-based system, Monte Carlo simulations with 500 runs and 150 time steps per run are used.
The two estimators presented before, \ac{SEBKF} and Sampling-based \ac{SEBKF}, are compared under the use of different weighting functions in the stochastic trigger.
The \ac{MMSE}-optimal estimation result obtained by a standard \ac{KF} that receives all sensor measurements periodically is plotted as a reference.
The implicit measurements \(\triggerconst_k\) are determined according to an \ac{SODP} policy with \(\triggerconst[k] = \measmat\sysmat^{l}\est^\text{S}_{k_e}\).
The Sampling-based \ac{SEBKF} is used with \(N=1000\) samples if not denoted otherwise.
As performance metrics, the \ac{MSE} relative to the \ac{MSE} of \ac{SEBKF} and the \ac{ANEES}~\cite{bar-shalomEstimationApplicationsTracking2001} are used to ensure low estimation errors and reliable estimation error covariance matrices.

\subsection{System Model}
As a system model, a nearly-constant velocity model in 2D is considered. The system model is characterized by the following matrices 
\begin{align}
&\sysmat[] = \left[\begin{smallmatrix}1 & \Delta & 0 & 0 \\ 0 & 1 & 0 & 0 \\ 0& 0& 1 & \Delta \\ 0&0&0&1\end{smallmatrix}\right] \,,
&\measmat[] = \begin{bmatrix}1 & 0 & 0 & 0 \\ 0&0 &1 & 0\end{bmatrix}\,, \end{align}\begin{align}
& \syscov[] = q \cdot \left[\begin{smallmatrix}\Delta^3/3 & \Delta^2/2 & 0&0 \\ \Delta^2/2 & \Delta &0&0 \\ 0&0&\Delta^3/3 & \Delta^2/2 \\ 0&0&\Delta^2/2 &\Delta\end{smallmatrix}\right] \,,
& \meascov[] = \begin{bmatrix}1&0\\0&1\end{bmatrix} \,,
\end{align}
where $\Delta=0.3$ is the sampling interval, and $q=1$ is a power scaling factor.

\begin{figure*}
    \setlength\fheight{0.94\figheight}
    \setlength\fwidth{\linewidth}
\begin{subfigure}{0.5\textwidth}\centering
%
%
\definecolor{mycolor1}{rgb}{0.00000,0.44700,0.74100}%
\definecolor{mycolor2}{rgb}{0.85000,0.32500,0.09800}%
\definecolor{mycolor3}{rgb}{0.92900,0.69400,0.12500}%
\definecolor{mycolor4}{rgb}{0.49400,0.18400,0.55600}%
\begin{tikzpicture}

\begin{axis}[%
width=\textwidth,
height=\fheight,
at={(0,0)},
xmin=0,
xmax=1,
xlabel style={font=\color{white!15!black}},
xlabel={$\bar{\gamma}$},
ymin=0,
ymax=1.2,
ylabel style={font=\color{white!15!black}},
ylabel={relative MSE},
axis background/.style={fill=white},
legend style={at={(axis cs:1,0)}, anchor=south east, legend cell align=left, align=left, draw=white!15!black}
]
\addplot [color=mycolor1, dashed]
  table[row sep=crcr]{%
0.994562814070348	1.00004074069717\\
0.946502512562813	0.99762636663359\\
0.898773869346735	0.991844021654586\\
0.78513567839196	0.965487708579754\\
0.652291457286432	0.90239164307776\\
0.558713567839195	0.847974911703254\\
0.500572864321608	0.799129972619919\\
0.417618090452262	0.717409004409011\\
0.361125628140703	0.651884632501182\\
0.322361809045226	0.610088051188237\\
0.274090452261307	0.532584638996317\\
0.229517587939699	0.457219177409481\\
0.166854271356784	0.324786686167173\\
0.1138391959799	0.201525200027713\\
0.0883316582914571	0.134066091397312\\
0.069979899497487	0.08569550632901\\
0.0622713567839192	0.0655052782976904\\
0.0527437185929646	0.0463040830956244\\
0.0440603015075374	0.0289559008555526\\
0.0289949748743717	0.00878857769224245\\
0.0247738693467336	0.00529990889403363\\
0.0206432160804019	0.0031004857518332\\
};
\addlegendentry{KF}


\addplot [color=mycolor3, dashed]
  table[row sep=crcr]{%
0.994562814070348	1\\
0.946502512562813	1\\
0.898773869346735	1\\
0.78513567839196	1\\
0.652291457286432	1\\
0.558713567839195	1\\
0.500572864321608	1\\
0.417618090452262	1\\
0.361125628140703	1\\
0.322361809045226	1\\
0.274090452261307	1\\
0.229517587939699	1\\
0.166854271356784	1\\
0.1138391959799	1\\
0.0883316582914571	1\\
0.069979899497487	1\\
0.0622713567839192	1\\
0.0527437185929646	1\\
0.0440603015075374	1\\
0.0289949748743717	1\\
0.0247738693467336	1\\
0.0206432160804019	1\\
};
\addlegendentry{SEBKF}

\addplot [color=mycolor4, mark=x, mark options={solid, mycolor4}]
  table[row sep=crcr]{%
0.994562814070348	1.00003719550773\\
0.946502512562813	1.00027267952402\\
0.898773869346735	1.00093145337136\\
0.78513567839196	1.00093024715658\\
0.652291457286432	1.00172431683624\\
0.558713567839195	1.00228203036277\\
0.500572864321608	1.00285602253379\\
0.417618090452262	1.00336938760383\\
0.361125628140703	1.00316074065626\\
0.322361809045226	1.00348139185266\\
0.274090452261307	1.00353381430974\\
0.229517587939699	1.00427005666869\\
0.166854271356784	1.00307420657218\\
0.1145              1.0044\\
0.0881              1.0045\\
0.0708              1.0064\\
0.0618              1.0038\\
0.0532              1.0036\\
0.0434              1.0057\\
0.0292              1.0025\\
0.0247              1.0106\\
0.0209              1.0023\\
};
\addlegendentry{Sample SEBKF}

\end{axis}
\end{tikzpicture}%
  \caption{Relative MSE over event rate for $\beta=2$.}
 \label{fig:relMSE2}
 \end{subfigure}
 \begin{subfigure}{0.5\textwidth}\centering
%
%
\definecolor{mycolor1}{rgb}{0.00000,0.44700,0.74100}%
\definecolor{mycolor2}{rgb}{0.85000,0.32500,0.09800}%
\definecolor{mycolor3}{rgb}{0.92900,0.69400,0.12500}%
\definecolor{mycolor4}{rgb}{0.49400,0.18400,0.55600}%
\begin{tikzpicture}

\begin{axis}[%
width=\textwidth,
height=\fheight,
at={(0,0)},
xmin=0,
xmax=1,
xlabel style={font=\color{white!15!black}},
xlabel={$\bar{\gamma}$},
ymin=0.7,
ymax=2,
ylabel style={font=\color{white!15!black}},
ylabel={ANEES},
axis background/.style={fill=white},
legend style={at={(axis cs:1,2)}, anchor=north east, legend cell align=left, align=left, draw=white!15!black}
]
\addplot [color=mycolor1, mark=x, mark options={solid, mycolor1}]
  table[row sep=crcr]{%
0.994562814070348	1.00301846279479\\
0.946502512562813	0.995012750650775\\
0.898773869346735	1.00504747679148\\
0.78513567839196	1.0058613350624\\
0.652291457286432	0.999802374074409\\
0.558713567839195	1.00092486672913\\
0.500572864321608	0.996848449033167\\
0.417618090452262	0.998826541331864\\
0.361125628140703	0.999497441886773\\
0.322361809045226	1.00072899982887\\
0.274090452261307	0.997242830478948\\
0.229517587939699	0.997151010000786\\
0.166854271356784	1.00141557395936\\
0.1138391959799	0.998824228892954\\
0.0883316582914571	0.995781515390769\\
0.069979899497487	0.999901308053406\\
0.0622713567839192	1.00135236452777\\
0.0527437185929646	1.00827980351658\\
0.0440603015075374	0.996880601467671\\
0.0289949748743717	0.994020485489809\\
0.0247738693467336	0.995951704884015\\
0.0206432160804019	1.00059841210855\\
};
\addlegendentry{KF}


\addplot [color=mycolor3, mark=x, mark options={solid, mycolor3}]
  table[row sep=crcr]{%
0.994562814070348	1.00297588663811\\
0.946502512562813	0.995118314936005\\
0.898773869346735	1.00506688432275\\
0.78513567839196	1.0037646361184\\
0.652291457286432	1.00218433436388\\
0.558713567839195	0.999022899033111\\
0.500572864321608	0.996206511393278\\
0.417618090452262	0.997791301099209\\
0.361125628140703	1.00093693142314\\
0.322361809045226	0.989538195631747\\
0.274090452261307	0.996624776297919\\
0.229517587939699	0.997000148059316\\
0.166854271356784	1.01201128261474\\
0.1138391959799	1.00116753804277\\
0.0883316582914571	0.992353354994083\\
0.069979899497487	0.995833306405802\\
0.0622713567839192	1.00459985103906\\
0.0527437185929646	0.996094380011548\\
0.0440603015075374	0.978816046713351\\
0.0289949748743717	0.971741535616381\\
0.0247738693467336	0.97983005853632\\
0.0206432160804019	0.983086832786\\
};
\addlegendentry{SEBKF}

\addplot [color=mycolor4, mark=x, mark options={solid, mycolor4}]
  table[row sep=crcr]{%
0.994562814070348	1.00323176452748\\
0.946502512562813	0.996262360716392\\
0.898773869346735	1.00679304361082\\
0.78513567839196	1.0069563357664\\
0.652291457286432	1.00701827943515\\
0.558713567839195	1.00410020930696\\
0.500572864321608	1.00126797413345\\
0.417618090452262	1.00394529546208\\
0.361125628140703	1.00601328890772\\
0.322361809045226	0.994451061651908\\
0.274090452261307	0.999805874472653\\
0.229517587939699	0.999460512639271\\
0.166854271356784	1.01214579191135\\
0.1145              0.9956\\
0.0881              1.0057\\
0.0708              0.9991\\
0.0618              0.9970\\
0.0532              1.0076\\
0.0434              1.0009\\
0.0292              0.9938\\
0.0247              1.0014\\
0.0209              0.9836\\
};
\addlegendentry{Sample SEBKF}

\end{axis}
\end{tikzpicture}%
  \caption{ANEES over event rate for $\beta=2$.}
 \label{fig:ANEES2}
 \end{subfigure}
 \vspace{0.5cm}
\begin{subfigure}{0.5\textwidth}\centering
%
%
\definecolor{mycolor1}{rgb}{0.00000,0.44700,0.74100}%
\definecolor{mycolor2}{rgb}{0.85000,0.32500,0.09800}%
\definecolor{mycolor3}{rgb}{0.92900,0.69400,0.12500}%
\definecolor{mycolor4}{rgb}{0.49400,0.18400,0.55600}%
\begin{tikzpicture}

\begin{axis}[%
width=\textwidth,
height=\fheight,
at={(0,0)},
xmin=0,
xmax=1,
xlabel style={font=\color{white!15!black}},
xlabel={$\bar{\gamma}$},
ymin=0,
ymax=1.2,
ylabel style={font=\color{white!15!black}},
ylabel={relative MSE},
axis background/.style={fill=white},
legend style={at={(axis cs:1,0)}, anchor=south east, legend cell align=left, align=left, draw=white!15!black}
]
\addplot [color=mycolor1, dashed]
  table[row sep=crcr]{%
0.996633165829142	0.999992936711969\\
0.96710552763819	0.999457249134744\\
0.935738693467335	0.998469601885511\\
0.849266331658291	0.98968256207939\\
0.730874371859297	0.961269703236927\\
0.6335175879397	0.919336716231862\\
0.552773869346733	0.877740300602025\\
0.441185929648242	0.800611252524785\\
0.363658291457286	0.727372685956491\\
0.312763819095477	0.671373407728277\\
0.242874371859297	0.575878769393008\\
0.190884422110553	0.480408944629101\\
0.12556783919598	0.314590997298283\\
0.0807437185929645	0.166368536421984\\
0.0622713567839192	0.0996088494095616\\
0.0495175879396983	0.0563298102194589\\
0.042582914572864	0.0400685335244678\\
0.0363819095477384	0.026452572468423\\
0.0295879396984923	0.0151133822138091\\
0.0191557788944722	0.00374358040240312\\
0.015788944723618	0.00196988913602459\\
0.013718592964824	0.00108624812893352\\
};
\addlegendentry{KF}

\addplot [color=mycolor3, dashed]
  table[row sep=crcr]{%
0.996633165829142	1\\
0.96710552763819	1\\
0.935738693467335	1\\
0.849266331658291	1\\
0.730874371859297	1\\
0.6335175879397	1\\
0.552773869346733	1\\
0.441185929648242	1\\
0.363658291457286	1\\
0.312763819095477	1\\
0.242874371859297	1\\
0.190884422110553	1\\
0.12556783919598	1\\
0.0807437185929645	1\\
0.0622713567839192	1\\
0.0495175879396983	1\\
0.042582914572864	1\\
0.0363819095477384	1\\
0.0295879396984923	1\\
0.0191557788944722	1\\
0.015788944723618	1\\
0.013718592964824	1\\
};
\addlegendentry{SEBKF}

\addplot [color=mycolor4, mark=x, mark options={solid, mycolor4}]
  table[row sep=crcr]{%
0.996633165829142	0.9999993148306\\
0.96710552763819	1.00013284606983\\
0.935738693467335	1.00008422032365\\
0.849266331658291	0.999819539967299\\
0.730874371859297	0.998288900600506\\
0.6335175879397	0.996415034098724\\
0.552773869346733	0.993671369131425\\
0.441185929648242	0.988670576414225\\
0.363658291457286	0.98625218594493\\
0.312763819095477	0.986505848305431\\
0.242874371859297	0.986652739902084\\
0.190884422110553	0.988333615558379\\
0.12556783919598	0.996507197709446\\
0.0807437185929645	1.00743810399031\\
0.0622713567839192	1.01481727879092\\
0.0495175879396983	1.0222493138122\\
0.042582914572864	1.02795193387104\\
0.0363819095477384	1.04437485225223\\
0.0295879396984923	1.04934762770161\\
0.0191557788944722	1.07879662198871\\
0.015788944723618	1.06943292207035\\
0.013718592964824	1.06730707890401\\
};
\addlegendentry{Sample SEBKF}

\end{axis}
\end{tikzpicture}%
  \caption{Relative MSE over event rate $\beta=5$.}
 \label{fig:relMSE30}
 \end{subfigure}
 \begin{subfigure}{0.5\textwidth}\centering
%
%
\definecolor{mycolor1}{rgb}{0.00000,0.44700,0.74100}%
\definecolor{mycolor2}{rgb}{0.85000,0.32500,0.09800}%
\definecolor{mycolor3}{rgb}{0.92900,0.69400,0.12500}%
\definecolor{mycolor4}{rgb}{0.49400,0.18400,0.55600}%
\begin{tikzpicture}

\begin{axis}[%
width=\textwidth,
height=\fheight,
at={(0,0)},
xmin=0,
xmax=1,
xlabel style={font=\color{white!15!black}},
xlabel={$\bar{\gamma}$},
ymin=0.7,
ymax=2,
ylabel style={font=\color{white!15!black}},
ylabel={ANEES},
axis background/.style={fill=white},
legend style={at={(axis cs:1,2)}, anchor=north east,legend cell align=left, align=left, draw=white!15!black}
]
\addplot [color=mycolor1, mark=x, mark options={solid, mycolor1}]
  table[row sep=crcr]{%
0.996633165829142	0.994794371010114\\
0.96710552763819	1.00252970952239\\
0.935738693467335	0.996802420287221\\
0.849266331658291	1.01140901894142\\
0.730874371859297	0.999988524733938\\
0.6335175879397	0.998495305718558\\
0.552773869346733	1.00026152069539\\
0.441185929648242	1.00006097526687\\
0.363658291457286	1.00278625219414\\
0.312763819095477	1.00601817486041\\
0.242874371859297	0.999579157214487\\
0.190884422110553	1.00559332963464\\
0.12556783919598	0.988446642769504\\
0.0807437185929645	0.999621507408439\\
0.0622713567839192	1.01187259611897\\
0.0495175879396983	1.00411372738601\\
0.042582914572864	0.995435718761205\\
0.0363819095477384	1.00251842138923\\
0.0295879396984923	1.0024022751303\\
0.0191557788944722	0.995602165097875\\
0.015788944723618	1.00141521775498\\
0.013718592964824	0.98973813491045\\
};
\addlegendentry{KF}

\addplot [color=mycolor3, mark=x, mark options={solid, mycolor3}]
  table[row sep=crcr]{%
0.996633165829142	0.994792564234501\\
0.96710552763819	1.00168817373688\\
0.935738693467335	0.993024993598683\\
0.849266331658291	0.995300073424245\\
0.730874371859297	0.964132856500585\\
0.6335175879397	0.955345312494809\\
0.552773869346733	0.940008990174064\\
0.441185929648242	0.932341530813686\\
0.363658291457286	0.933275838886017\\
0.312763819095477	0.928858908860642\\
0.242874371859297	0.927292971560307\\
0.190884422110553	0.938921206877684\\
0.12556783919598	0.958089614879559\\
0.0807437185929645	1.01866658182854\\
0.0622713567839192	1.05085791941861\\
0.0495175879396983	1.1112503013071\\
0.042582914572864	1.11567735050907\\
0.0363819095477384	1.14486072848752\\
0.0295879396984923	1.14906309615891\\
0.0191557788944722	1.25598582487451\\
0.015788944723618	1.36380368903738\\
0.013718592964824	1.39596660177948\\
};
\addlegendentry{SEBKF}

\addplot [color=mycolor4, mark=x, mark options={solid, mycolor4}]
  table[row sep=crcr]{%
0.996633165829142	0.994824510684399\\
0.96710552763819	1.00294699656952\\
0.935738693467335	0.996909374318689\\
0.849266331658291	1.01080869887477\\
0.730874371859297	0.999568960700029\\
0.6335175879397	1.00692442679125\\
0.552773869346733	1.00353783187253\\
0.441185929648242	1.00579591585327\\
0.363658291457286	1.01174134663857\\
0.312763819095477	1.0078528241184\\
0.242874371859297	0.999869284859326\\
0.190884422110553	0.999400369955945\\
0.12556783919598	0.985626242678943\\
0.0807437185929645	0.990590997664857\\
0.0622713567839192	0.987095864292198\\
0.0495175879396983	1.01158392521296\\
0.042582914572864	0.992237571412499\\
0.0363819095477384	1.0037036382992\\
0.0295879396984923	0.973533487490787\\
0.0191557788944722	0.977371094666315\\
0.015788944723618	0.995543839908579\\
0.013718592964824	0.986738202674685\\
};
\addlegendentry{Sample SEBKF}

\end{axis}
\end{tikzpicture}%
  \caption{ANEES over event rate for $\beta=5$.}
 \label{fig:ANEES30}
 \end{subfigure}
  \vspace{0.5cm}
\begin{subfigure}{0.5\textwidth}\centering
%
%
\definecolor{mycolor1}{rgb}{0.00000,0.44700,0.74100}%
\definecolor{mycolor2}{rgb}{0.85000,0.32500,0.09800}%
\definecolor{mycolor3}{rgb}{0.92900,0.69400,0.12500}%
\definecolor{mycolor4}{rgb}{0.49400,0.18400,0.55600}%
\begin{tikzpicture}

\begin{axis}[%
width=\textwidth,
height=\fheight,
at={(0,0)},
xmin=0,
xmax=1,
xlabel style={font=\color{white!15!black}},
xlabel={$\bar{\gamma}$},
ymin=0,
ymax=1.2,
ylabel style={font=\color{white!15!black}},
ylabel={relative MSE},
axis background/.style={fill=white},
legend style={at={(axis cs:1,0)}, anchor=south east, legend cell align=left, align=left, draw=white!15!black}
]
\addplot [color=mycolor1, dashed]
  table[row sep=crcr]{%
0.99747738693467	0.999996679990022\\
0.97177889447236	0.999719054274086\\
0.944201005025124	0.998429970698937\\
0.869326633165829	0.991706284286149\\
0.757638190954774	0.97054462709391\\
0.661216080402011	0.943057368643397\\
0.58129648241206	0.91000647936642\\
0.460834170854272	0.83774813582972\\
0.371758793969849	0.761458785452584\\
0.313577889447236	0.707550858777622\\
0.237668341708543	0.599059887418552\\
0.17721608040201	0.493725186764142\\
0.113628140703518	0.308969503079856\\
0.0711758793969845	0.150014031587743\\
0.0536984924623113	0.0840767371923445\\
0.0416984924623112	0.0441167916016905\\
0.0357286432160801	0.0309460445248332\\
0.030542713567839	0.0189093153278421\\
0.0245326633165828	0.00964149947377982\\
0.0151758793969848	0.00198660812566397\\
0.0126432160804019	0.00101761629334616\\
0.010572864321608	0.000547200967563233\\
};
\addlegendentry{KF}


\addplot [color=mycolor3, dashed]
  table[row sep=crcr]{%
0.99747738693467	1\\
0.97177889447236	1\\
0.944201005025124	1\\
0.869326633165829	1\\
0.757638190954774	1\\
0.661216080402011	1\\
0.58129648241206	1\\
0.460834170854272	1\\
0.371758793969849	1\\
0.313577889447236	1\\
0.237668341708543	1\\
0.17721608040201	1\\
0.113628140703518	1\\
0.0711758793969845	1\\
0.0536984924623113	1\\
0.0416984924623112	1\\
0.0357286432160801	1\\
0.030542713567839	1\\
0.0245326633165828	1\\
0.0151758793969848	1\\
0.0126432160804019	1\\
0.010572864321608	1\\
};
\addlegendentry{SEBKF}

\addplot [color=mycolor4, mark=x, mark options={solid, mycolor4}]
  table[row sep=crcr]{%
0.99747738693467	1.00003274031757\\
0.97177889447236	1.00022099483238\\
0.944201005025124	1.00007901568621\\
0.869326633165829	1.00025945730329\\
0.757638190954774	0.99917817355204\\
0.661216080402011	0.997026261187\\
0.58129648241206	0.993725375994522\\
0.460834170854272	0.986267139204648\\
0.371758793969849	0.980755531662768\\
0.313577889447236	0.979746752758842\\
0.237668341708543	0.978047742106763\\
0.17721608040201	0.982029900989749\\
0.113628140703518	0.998237747732358\\
0.0710              1.0065\\
0.0542              1.0153\\
0.0418              1.0367\\
0.0353              1.0596\\
0.0307              1.0636\\
0.0243              1.0666\\
0.0153              1.0784\\
0.0126              1.0710\\
0.0108              1.0336\\
};
\addlegendentry{Sample SEBKF}

\end{axis}

\end{tikzpicture}%
  \caption{Relative MSE over event rate $\beta=1000$.}
 \label{fig:relMSE1000}
 \end{subfigure}
 \begin{subfigure}{0.5\textwidth}\centering
%
%
\definecolor{mycolor1}{rgb}{0.00000,0.44700,0.74100}%
\definecolor{mycolor2}{rgb}{0.85000,0.32500,0.09800}%
\definecolor{mycolor3}{rgb}{0.92900,0.69400,0.12500}%
\definecolor{mycolor4}{rgb}{0.49400,0.18400,0.55600}%
\begin{tikzpicture}

\begin{axis}[%
width=\textwidth,
height=\fheight,
at={(0,0)},
xmin=0,
xmax=1,
xlabel style={font=\color{white!15!black}},
xlabel={$\bar{\gamma}$},
ymin=0.7,
ymax=2,
ylabel style={font=\color{white!15!black}},
ylabel={ANEES},
axis background/.style={fill=white},
legend style={at={(axis cs:1,2)}, anchor=north east,legend cell align=left, align=left, draw=white!15!black}
]
\addplot [color=mycolor1, mark=x, mark options={solid, mycolor1}]
  table[row sep=crcr]{%
0.99747738693467	0.99895501658654\\
0.97177889447236	0.997682170939329\\
0.944201005025124	0.990918251317707\\
0.869326633165829	0.998223088614151\\
0.757638190954774	0.993115567078048\\
0.661216080402011	0.997980895327149\\
0.58129648241206	0.993973101443854\\
0.460834170854272	1.00401742693968\\
0.371758793969849	0.996403995653398\\
0.313577889447236	0.997242672808241\\
0.237668341708543	0.994368123320979\\
0.17721608040201	1.00476025501378\\
0.113628140703518	0.999248352951565\\
0.0711758793969845	0.989121542977374\\
0.0536984924623113	1.00595869795955\\
0.0416984924623112	0.997096132834155\\
0.0357286432160801	1.00057834805405\\
0.030542713567839	0.992002351099904\\
0.0245326633165828	0.998546973360698\\
0.0151758793969848	0.997524312894474\\
0.0126432160804019	0.995453996341565\\
0.010572864321608	0.99908635339795\\
};
\addlegendentry{KF}


\addplot [color=mycolor3, mark=x, mark options={solid, mycolor3}]
  table[row sep=crcr]{%
0.99747738693467	0.998948999321878\\
0.97177889447236	0.996656634849773\\
0.944201005025124	0.987666389488264\\
0.869326633165829	0.983422057643644\\
0.757638190954774	0.956361919363105\\
0.661216080402011	0.939844310244122\\
0.58129648241206	0.921481224366408\\
0.460834170854272	0.914331792745646\\
0.371758793969849	0.900884418863048\\
0.313577889447236	0.89656732873449\\
0.237668341708543	0.901309979484026\\
0.17721608040201	0.91347117371976\\
0.113628140703518	0.966891551968512\\
0.0711758793969845	1.05808988479568\\
0.0536984924623113	1.13239178280507\\
0.0416984924623112	1.22844198253863\\
0.0357286432160801	1.25899857071215\\
0.030542713567839	1.34904234570338\\
0.0245326633165828	1.51219500157395\\
0.0151758793969848	1.94283917111028\\
0.0126432160804019	2.12427606191028\\
0.010572864321608	2.24256681288158\\
};
\addlegendentry{SEBKF}

\addplot [color=mycolor4, mark=x, mark options={solid, mycolor4}]
  table[row sep=crcr]{%
0.99747738693467	0.999000007221998\\
0.97177889447236	0.998405216167859\\
0.944201005025124	0.992064749736817\\
0.869326633165829	1.00062044242647\\
0.757638190954774	0.997207694132341\\
0.661216080402011	1.00236894751972\\
0.58129648241206	0.996250300779098\\
0.460834170854272	1.00721393575292\\
0.371758793969849	1.00347490378907\\
0.313577889447236	0.99741116586235\\
0.237668341708543	0.999416335030342\\
0.17721608040201	0.993948332233198\\
0.113628140703518	1.0006688454408\\
0.0710              0.9809\\
0.0542              1.0019\\
0.0418              0.9961\\
0.0353              1.0258\\
0.0307              1.0362\\
0.0243              1.0093\\
0.0153              1.0166\\
0.0126              1.0205\\
0.0108              0.9630\\
};
\addlegendentry{Sample SEBKF}

\end{axis}
\end{tikzpicture}%
  \caption{ANEES over event rate for $\beta=1000$.}
 \label{fig:ANEES1000}
 \end{subfigure}
 \caption{System performance under the use of Gaussian stochastic triggers using different values of $\beta$ and different estimators for state estimation at the receiver.}
 \label{fig:SimGaussianWeightingFcts}
 \end{figure*}
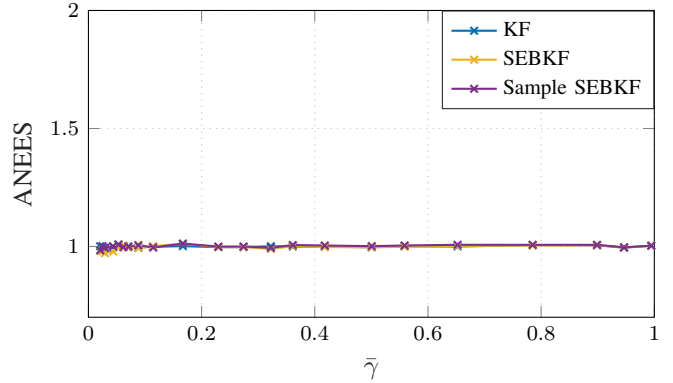
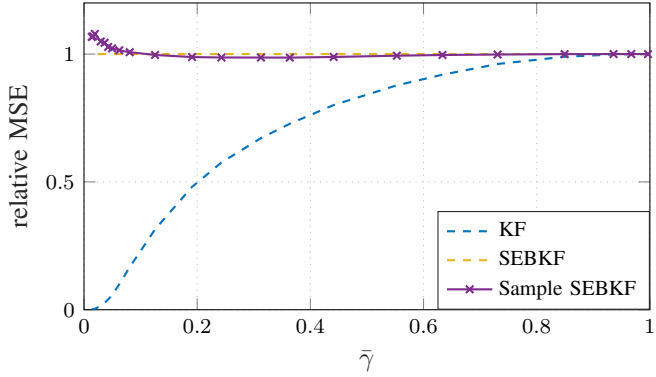
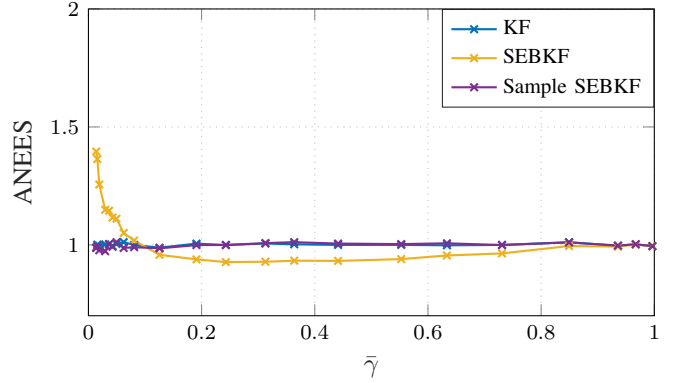
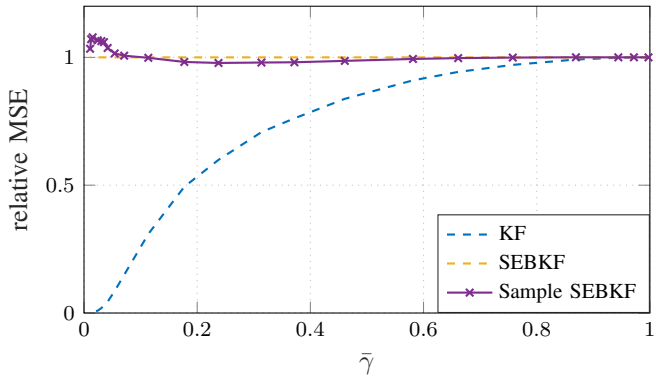
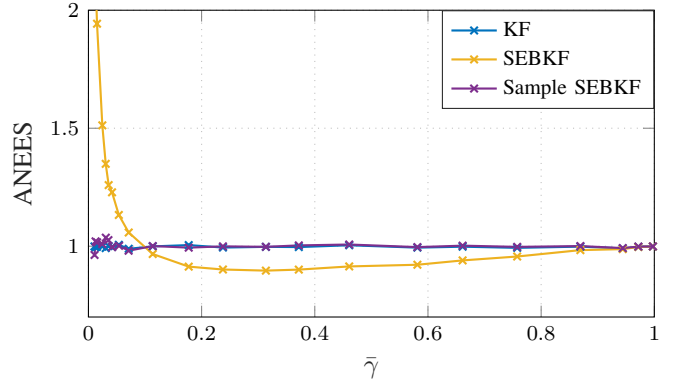

\subsection{Gaussian Weighting Functions}
The Gaussian weighting functions \(\phi_\beta({\rvec{z}}_k)\) are evaluated for \(\beta=2,\,5,\,1000\), where \(\beta=1000\) closely resembles the deterministic triggering case under the previously defined conditions.
In Fig.~\ref{fig:SimGaussianWeightingFcts}, the \ac{MSE} relative to \ac{SEBKF} and the \ac{ANEES} of each estimator are shown for the three values of \(\beta\). It can be observed that for \(\beta=2\), \ac{SEBKF} and Sampling-based \ac{SEBKF} show the same (optimal) performance regarding the \ac{MSE} and are efficient (\ac{ANEES} \(\approx 1\)) for all event rates. This means that for \(\beta=2\) the Sampling-based \ac{SEBKF} asymptotically reaches the optimal result if sufficient samples are used. For \(\beta=5\) and \(\beta=1000\) \ac{SEBKF} is slightly conservative for medium event rates and becomes inconsistent (\ac{ANEES} \(> 1\)) for low event rates \(\approx<10\%\), which was expected from Fig.~\ref{fig:EventRateZConstVel} and Fig.~\ref{fig:zZz_ZConstVel}. The Sampling-based \ac{SEBKF} remains consistent beyond this point if the number of samples is increased sufficiently, here \(N=5000\) samples are used. Performance-wise Sampling-based \ac{SEBKF} has a similar performance as \ac{SEBKF}, however, the performance starts to decrease for very low event rates.

It can be concluded that \ac{SEBKF} behaves as expected with weighting functions \(\beta>2\) and therefore can only be used safely for high enough transmission rates. Sampling-based \ac{SEBKF} can be used for all transmission rates and provides similar estimation results as \ac{SEBKF}, but it comes at the cost of higher computational load and storage requirements at the remote receiver.

\section{Conclusions}
In this work, a generalized framework for stochastic innovation-based triggering was proposed. First, deterministic triggering policies were fitted into the stochastic framework by the introduction of generalized Gaussian weighting functions for stochastic triggers.
Furthermore, criteria were developed to evaluate the relative achievable performance of the remote estimator considering only the system model and the design of the stochastic trigger. We then evaluated under which conditions the \ac{SEBKF} estimator can be used safely with Gaussian weighting functions other than \(\phi_2(\rvec{z}_k)\).
An event-based particle filtering approach was extended to stochastic triggers and improved for the use with linear \acp{SSM}. Moreover, the performance of different combinations of weighting functions and estimators was evaluated in simulations, especially regarding consistency using the \ac{ANEES}. The results obtained in the simulation studies confirmed the considerations made regarding the applicability of the \ac{SEBKF} under different conditions and showed the strengths and weaknesses of the sampling-based estimator.

In future, a formal analysis of the functional connection between the event rate, the choice \(\triggermat\) and the error introduced by the triggering function for given weighting functions and system models would be desirable. Furthermore, strict conditions for the consistency of \ac{SEBKF} should be established. Additionally, further generalization of the stochastic trigger to other possible weighting functions together with potential use cases should be investigated, also considering extensions to nonlinear system models. Moreover, the sampling-based approach presented in this work might prove beneficial in situations, in which the Gaussian property of the state variable cannot be maintained, e.\,g., if packet losses are present in the communications system or sensor measurements are not available at all times~\cite{kungNonexistenceEventBasedTriggers2017, MFI24_Schmitt}.

\section*{Acknowledgment}
This work was partially funded by the Deutsche Forschungsgemeinschaft (DFG, German Research Foundation) under project number 515674308.

\balance
\bibliographystyle{IEEEtran}
\bibliography{BibTeX/ieeeBibControls,BibTeX/ams,BibTeX/EBE,BibTeX/EBE2,BibTeX/T2TF,BibTeX/other}

\end{document}